\documentclass[a4paper,11pt]{article}
\usepackage{pos}
\usepackage{amsmath,leftidx}
\usepackage{dsfont}
\usepackage{bbm}
\usepackage{cleveref}
\usepackage{subcaption}
\usepackage[utf8]{inputenc}
\usepackage{mathrsfs}
\usepackage{amsfonts}
\usepackage{wrapfig}
\usepackage{enumitem}
\usepackage{booktabs}
\usepackage{tikz}
\usepackage{physics}

\usetikzlibrary{decorations.pathreplacing,calc,tikzmark,patterns,fit,positioning}

\title{The hadronic contribution to the running of $\alpha$ and the electroweak mixing angle}

\author*[a,b]{Alessandro Conigli}
\author[c]{Georg von Hippel}
\author[d]{Simon Kuberski}
\author[a,b,c]{Harvey B. Meyer}
\author[c]{Konstantin Ottnad}
\author[a,b,c]{Hartmut Wittig}

\affiliation[a]{Helmholtz Institute Mainz, Johannes Gutenberg-Universit\"{a}t Mainz, 55099 Mainz, Germany}
\affiliation[b]{GSI Helmholtz Centre for Heavy Ion Research, 64291 Darmstadt, Germany}
\affiliation[c]{PRISM$\mathrm{A}^+$ Cluster of Excellence and Institut f\"{u}r Kernphysik, Johannes Gutenberg-Universit\"{a}t Mainz, 55099 Mainz, Germany}
\affiliation[d]{Theoretical Physics Department, CERN, 1211 Geneva 23, Switzerland}

\emailAdd{aconigli@uni-mainz.de}

\abstract{
We report on our update to \cite{Ce:2022eix} on the hadronic running of electroweak couplings from $O(a)$-improved Wilson fermions with $N_f=2+1$ flavours. The inclusion of additional ensembles at very fine lattice spacings together with a number of techniques to split the different contributions for a better control of cutoff effects allows us to substantially improve the precision. We employ two different discretizations of the vector current to compute the subtracted Hadronic Vacuum Polarization (HVP) functions $\bar{\mathit{\Pi}}^{\gamma\gamma}$
and $\bar{\mathit{\Pi}}^{Z\gamma}$
for Euclidean time momenta up to $Q^2\leq 9 \ \mathrm{GeV}^2$ . To reduce cutoff effects in the short distance region we apply a suitable subtraction to the TMR kernel function, which cancels the leading $x_0^4$
behaviour. The subtracted term is then computed in perturbative QCD using the Adler function and added back to compensate for the subtraction. Chiral-continuum extrapolations are performed with five values of the lattice spacing and several pion masses, including its physical value, and several fit ansätze are explored to estimate the systematics arising from model selection. Our results show excellent prospects for high-precision estimates of 
$\Delta\alpha_{\mathrm{had}}^{(5)}(M_Z^2)$ at the Z-pole.

\begin{flushright}
	MITP-25-007 \\
	CERN-TH-2025-022
\end{flushright}
}

\FullConference{The 41st International Symposium on Lattice Field Theory (LATTICE2024)\\
 28 July - 3 August 2024\\
Liverpool, UK\\}

\begin{document}
\maketitle

\section{Introduction}
Precision observables play a crucial role in the search for new physics  beyond the Standard Model (SM). Complementary to direct searches  conducted at high-energy colliders,  theoretical predictions allow us to put constraints on  the  SM  by demanding that both theory and experimental measurements reach high precision. This can be especially challenging when the observable in question is influenced by significant hadronic effects. 
A notable example is the muon's anomalous magnetic moment $a_\mu$, where sub-percent precision was achieved both at the experimental   \cite{Muong-2:2021ojo,Muong-2:2023cdq} and theoretical level \cite{ Borsanyi:2020mff,Djukanovic:2024cmq, Boccaletti:2024guq, RBC:2024fic, Bazavov:2024eou}. In this context, Lattice QCD has emerged as one of the leading methods, replacing the data-driven approach with a first-principles calculation. 

Here we examine two closely related quantities that play a crucial role in SM tests: the energy dependence of the electromagnetic coupling $\alpha$ and the electroweak mixing angle $\sin^2\theta_W$. The former relates the square of the electric charge in the Thomson limit to its value at the Z-pole, which is an important input quantity in electroweak precision tests. Moreover, the running of the electroweak mixing angle is a sensitive probe of beyond-SM physics, particularly at low energies \cite{Becker:2018ggl}. As in the case of $a_\mu$, the overall precision of both the running of $\alpha$ and $sin^2\theta_W$ is limited by hadronic uncertainties.

\section{Methodology}

\subsection{Lattice setup}
We perform our calculation  on  a set of 26 $N_f=2+1$  ensembles generated by the Coordinated Lattice Simulations (CLS)   \cite{Bruno:2016plf,  Mohler:2017wnb, Mohler:2020txx, Kuberski:2023zky, Bali:2016umi} with non-perturbatively $O(a)$-improved Wilson fermions and a tree-level improved L\"{u}scher-Weisz gauge action. The gauge ensembles employed in this work cover five values of the lattice spacing in the range $0.0039\ \mathrm{fm} < a < 0.087 \ \mathrm{fm}$ while the pion masses lie in the range $130\ \mathrm{MeV}\leq m_\pi \leq 420 \ \mathrm{MeV}$.  We work with a subset of the CLS ensembles where the sum of the bare quark masses is kept constant when approaching the physical point, such that the bare coupling $\tilde{g}_0$ is also held constant along the chiral trajectory. To account for a small mistuning in $m_K^{\mathrm{phys}}$ when hitting $m_\pi^{\mathrm{phys}}$ we include four additional ensembles on a different chiral trajectory where the strange quark mass is fixed to its physical value.

To reduce cutoff effects, we perform a full $O(a)$ improvement of the observables with two independent sets of  non-perturbatively determined improvement coefficients:  by set 1 we denote  the results from large-volume simulations  as computed in \cite{Gerardin:2018kpy}, while we refer to set 2 when using $Z_V$ and $c_V$ from \cite{Heitger:2020zaq}  and $b_V,\bar{b}_V$  \cite{Fritzsch:2018zym}, determined in the Schr\"{o}dinger Functional (SF) setup. In addition, to further constrain the chiral-continuum extrapolations we employ two discretizations of the vector current, the local $(\mathit{L})$ and the point-split conserved $(\mathit{C})$ currents as defined in  \cite{Ce:2022kxy}.

\subsection{Time Momentum Representation}

The running of the electromagnetic coupling and the weak mixing angle at any momentum transfer $Q^2$ can be expressed in terms of the effective couplings as
\begin{equation}
	\alpha(Q^2) = \frac{\alpha}{1-\Delta\alpha(Q^2)}, \qquad
	\sin^2\theta_W(Q^2) = \sin^2\theta_W \big( 1 + \Delta\sin^2\theta_W(Q^2)\big),
\end{equation}
where $\alpha$ and $\sin^2\theta_W$ represent their values in the Thomson limit $Q^2=0$. The running $\Delta\alpha(Q^2)$ and $\Delta\sin^2\theta_W(Q^2)$ receive contributions from both the hadronic and leptonic sectors. While the latter can be computed reliably in perturbation theory, the contribution arising from low-energy quarks is non-perturbative and at leading order it can be expressed in terms of the subtracted HVP functions $\mathit{\bar{\Pi}}^{(\gamma,\gamma)}$ and $\mathit{\bar{\Pi}}^{(Z,\gamma)}$ as 
\begin{equation}
	\Delta\alpha_{\mathrm{had}}(Q^2) = 4\pi\alpha \mathit{\bar{\Pi}}^{(\gamma,\gamma)}(Q^2), \qquad
	(\Delta\sin^2\theta_W)_{\mathrm{had}}(Q^2) = -\frac{4\pi\alpha}{\sin^2\theta_W}\mathit{\bar{\Pi}}^{(Z,\gamma)}(Q^2),
\end{equation}
where $\mathit{\bar{\Pi}}(Q^2) = \mathit{\Pi}(Q^2) - \mathit{\Pi}(0)$. The terms  $\mathit{\bar{\Pi}}^{(\gamma,\gamma)}$ and $\mathit{\bar{\Pi}}^{(Z,\gamma)}$  can be computed on the lattice for any space-like momentum transfer through the Time-Momentum Representation (TMR) \cite{Bernecker:2011gh, Francis:2013fzp}, given by the integral over Euclidean time
\begin{equation}
	\mathit{\bar{\Pi}}^{(\alpha,\gamma)}(Q^2) = \int_{0}^{\infty} \dd{x_0}G^{(\alpha,\gamma)}(t)K(x_0, Q^2),
	\qquad \alpha = Z, \gamma,
	\label{eq:tmr_integral}
\end{equation}
where $G^{(\alpha,\gamma)}$ denotes the zero-momentum-projected vector correlator, while $K(x_0,Q^2)$ is a $Q^2$-dependent kernel function. Following the notation of \cite{Gerardin:2018kpy, Ce:2022eix} and working in the $SU(3)$-flavor basis, the vector correlators of interest read
\begin{eqnarray}
	G^{(\gamma,\gamma)} = G^{(3,3)} + \frac{1}{3}G^{(8,8)} + \frac{4}{9}G^{(c,c)} + \frac{4}{9}G^{(c,c)}_{\mathrm{disc}} + 
	\frac{2}{3\sqrt{3}}G^{(c,8)}_{\mathrm{disc}} + \ldots,
	\\
	G^{(Z,\gamma)} = \bigg(
	\frac{1}{2} - \sin^2\theta_W
	\bigg)G^{(\gamma,\gamma)} - \frac{1}{6\sqrt{3}}G^{(0,8)} - \frac{1}{18}G^{(c,c)} - \frac{1}{18}G^{(c,c)}_{\mathrm{disc}} + \ldots,
\end{eqnarray}
in terms of the building blocks as defined in \cite{Ce:2022eix}, which can be computed separately.

\subsection{Computational strategy}
To precisely  determine the value of $\Delta\alpha_{\mathrm{had}}^{(5)}(M_Z^2)$ at the Z-pole it is essential to reach high values of the momentum transfer $Q^2$ on the lattice. This ensures that the threshold energy above which the perturbative running becomes relevant is effectively shifted to higher values, therefore reducing the uncertainty from perturbation theory. However, achieving such  high $Q^2$ values  while maintaining control over the systematics involved in the computation  presents significant challenges. To address this, we propose the following decomposition for the subtracted HVP
\begin{equation}
	\mathit{\bar{\Pi}}(Q^2) = \big[
	\mathit{\Pi}(Q^2) -  \mathit{\Pi}(Q^2/4)
	\big] + \big[
	\mathit{\Pi}(Q^2/4) -  \mathit{\Pi}(0)
	\big].
	\label{eq:hvp_splitting}
\end{equation}
This facilitates a clear separation of contributions arising from different Euclidean distance scales,  thereby improving the control over the chiral-continuum extrapolations associated with each term. In the analysis presented here, we focus exclusively on the first term  on the right hand side of Eq.~(\ref{eq:hvp_splitting}), corresponding to  the high energy observable $\mathit{\widehat{\Pi}}(Q^2) =	\mathit{\Pi}(Q^2) -  \mathit{\Pi}(Q^2/4) $. 

\begin{figure}
	\centering
		\includegraphics[scale=0.44]{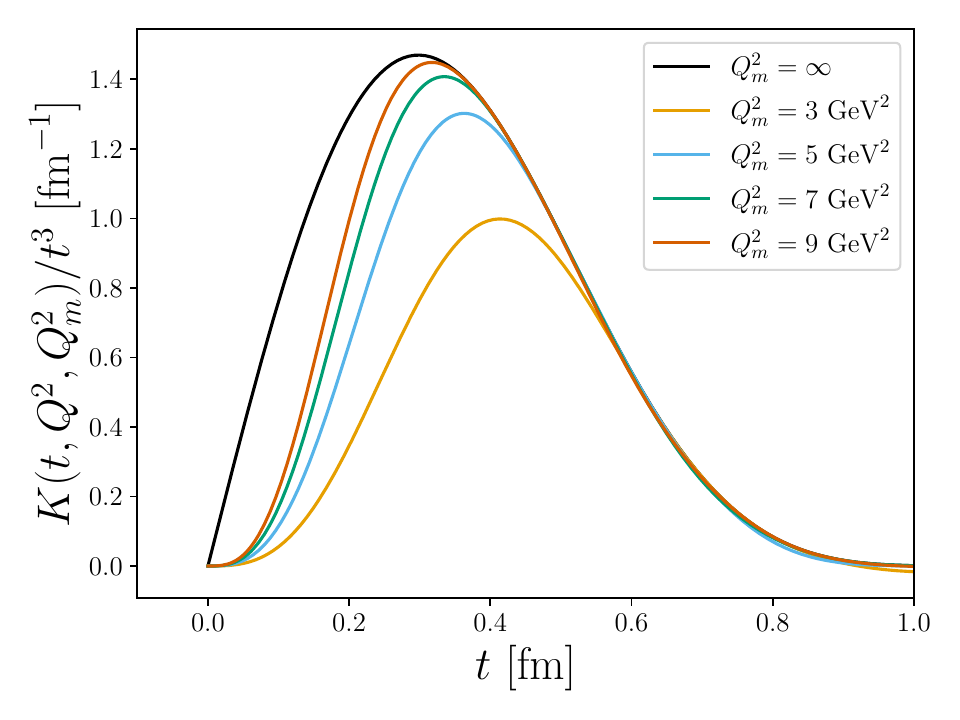}
	\includegraphics[scale=0.44]{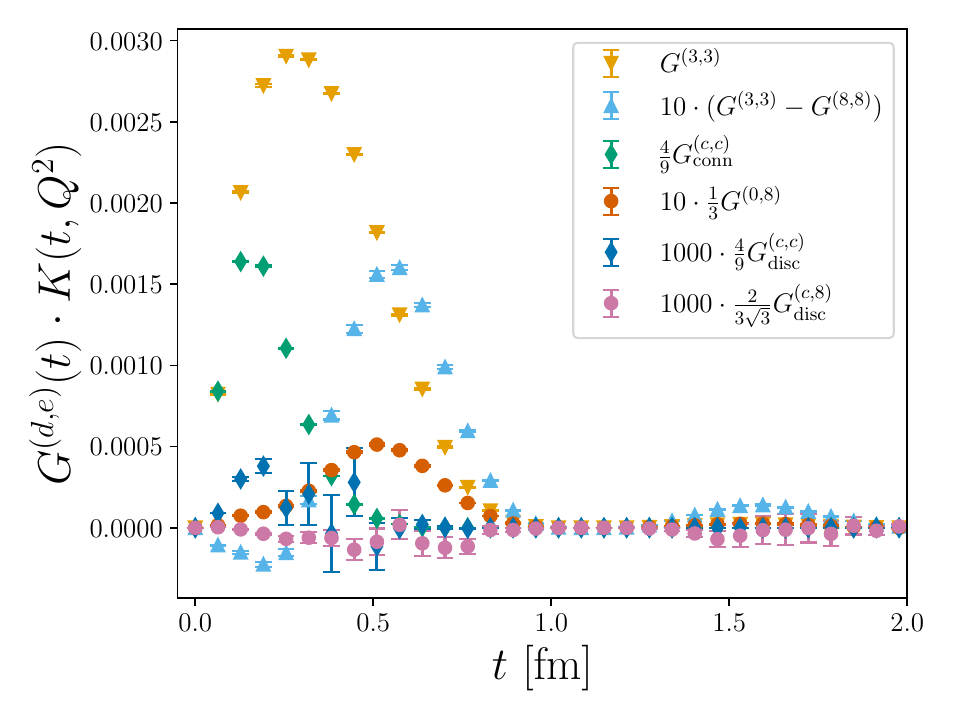}
	\caption{\textit{Left:} Illustration of the subtracted kernel in Eq.~(\ref{eq:subtracted_kernel}) for different values of the virtualities $Q_m$.  \textit{Right:} integrands of the various contributions according to Eq.~(\ref{eq:non_subtracted_kernel})  for $Q^2=5\ \mathrm{GeV}^2$. Results are shown for physical point ensemble E250 with $a \approx 0.064\ \mathrm{fm}$. }
	\label{fig:kernel_comparison}
\end{figure}

\paragraph{Isovector contribution:} following the approach detailed in \cite{Kuberski:2024bcj},  we extract the isovector contribution $\mathit{\widehat{\Pi}}^{(3,3)}(Q^2)$ using  the  decomposition
\begin{equation}
	\mathit{\widehat{\Pi}}^{(3,3)}(Q^2)
	=
	\mathit{\widehat{\Pi}}^{(3,3)}_{\mathrm{sub}}(Q^2) + b^{(3,3)}(Q^2, Q_m^2), 
\end{equation}
where $	\mathit{\widehat{\Pi}}^{(3,3)}_{\mathrm{sub}}(Q^2)$ is computed from the TMR integral Eq.~(\ref{eq:tmr_integral})  with the kernel function defined as
\begin{equation}
K_{\mathrm{sub}}(x_0, Q^2, Q_m^2) = \frac{16}{Q^2}\sin[4](\frac{Qx_0}{4})
-
\frac{Q^2}{Q_m^4}\sin[4](\frac{Q_mx_0}{2}).
\label{eq:subtracted_kernel}
\end{equation}
The second term,
\begin{equation}\label{eq:b33_def}
	b^{(3,3)}(Q^2, Q_m^2) = \frac{Q^2}{4Q_m^2} \bigg(
	\mathit{\Pi}^{(3,3)}(4Q_m^2) - \mathit{\Pi}^{(3,3)}(Q_m^2)
	\bigg),
\end{equation}
can be computed reliably using the massless perturbative Adler function.  The  subtracted kernel $K_{\mathrm{sub}}(x_0,Q^2, Q_m^2)$ is designed to cancel the $x_0^4$ behaviour in the TMR integral. The neat effect  is a reduction of higher-order cutoff effects, including  the potentially dangerous contributions such as $O(a^2\log(a))$ in the very short  Euclidean distance regime \cite{Ce:2021xgd, Sommer:2022wac}. As illustrated on the left-hand side of Fig.~\ref{fig:kernel_comparison}, the subtracted kernel function is shown for various choices of the virtualities $Q_m$. Throughout our analysis, we adopt  $Q_m = 3 \ \mathrm{GeV}$ as the default value unless otherwise specified.
Further reduction of  lattice artefacts originating from the Short-Distance (SD) region is  achieved  by incorporating tree-level perturbative corrections. Specifically, cutoff effects  are computed at tree-level in the massless theory, and the non-perturbatively evaluated observables are replaced with
\begin{equation}
	\mathcal{O}(a) \rightarrow \mathcal{O}(a) \frac{\mathcal{O}^{\mathrm{tl}}(0)}{\mathcal{O}^{\mathrm{tl}}(a)},
\end{equation} 
where  $\mathcal{O}^{\mathrm{tl}}(a)$ represents the tree-level evaluation of the observable. At our coarsest lattice spacing, this approach leads to a significant reduction of cutoff effects, from  approximately $20\%$  to $7\%$.

\paragraph{Isoscalar contribution:} after computing $\widehat{\mathit{\Pi}}^{(3,3)}$, the isoscalar contribution can be determined through the  decomposition \cite{Kuberski:2024bcj}
\begin{equation}
	\widehat{\mathit{\Pi}}^{(8,8)}(Q^2) = \widehat{\mathit{\Pi}}^{(3,3)}(Q^2) + \Delta_{ls}(Q^2),
	\label{eq:isoscalar_decomposition}
\end{equation}
where only the term $\Delta_{ls}$ has to be computed non-perturbatively. Notably,  $\Delta_{ls}$ is parametrically suppressed at short distances, therefore no assistance from perturbation theory is required to compute this observable. In practice, we evaluate $\Delta_{ls}$ from 
\begin{equation}
	\Delta_{ls}(Q^2) = \int_{0}^{\infty} \dd x_0 \big[
	G^{(8,8)} - G^{(3,3)} 
	\big] K(x_0, Q^2), \qquad 
	K(x_0, Q^2) = \frac{16}{Q^2}\sin[4](\frac{Qx_0}{4}),
	\label{eq:non_subtracted_kernel}
\end{equation} 
using the non-subtracted kernel function $K(x_0, Q^2)$.
A similar strategy is employed to compute the $\widehat{\mathit{\Pi}}^{(0,8)}$ contribution, which is relevant exclusively to the electroweak mixing angle.  This term vanishes linearly towards the $SU(3)$-symmetric point, ensuring that no additional perturbative calculations are required. Specifically, we fit the only available discretization (conserved-local) that avoids reliance on the renormalised singlet local current.

\paragraph{Charm connected contribution:} similar to the isovector case, the charm connected contribution can be extracted using the subtracted kernel Eq.~(\ref{eq:subtracted_kernel}), employing  the following decomposition
\begin{equation}
	\widehat{\mathit{\Pi}}^{(c,c)}(Q^2) = \widehat{\mathit{\Pi}}^{(c,c)}_{\mathrm{sub}}(Q^2) + 
	 b_{\mathrm{conn}}^{(c,c)}(Q^2, Q_m^2).
	 \label{eq:charm_decomposition}
\end{equation}
The subtraction function $ b_{\mathrm{conn}}^{(c,c)}$, defined analogously to $b^{(3,3)}$ in Eq.~(\ref{eq:b33_def}), is computed according to
\begin{equation}
	 b_{\mathrm{conn}}^{(c,c)}(Q^2, Q_m^2) = 2b^{(3,3)}(Q^2, Q_m^2) + \Delta_{lc}b(Q^2, Q_m^2),
\end{equation}
where the first term, $2b^{(3,3)}(Q^2, Q_m^2)$,  is derived from massless perturbation theory. The second term, $\Delta_{lc}b$, represents the difference between the isovector and charm-connected contributions, which we evaluate non-perturbatively. In the right panel of Fig.~\ref{fig:kernel_comparison}  we illustrate the behaviour and the relative size of the various contributions to the HVP.

\section{Chiral-continuum extrapolations}
Having computed all the relevant contributions across  five distinct  lattice spacings and multiple quark masses, we proceed with a  reliable extrapolation to the continuum and physical point. The latter is defined within the isospin-symmetric limit by fixing $m_\pi = (m_{\pi^0})_{\mathrm{phys}}$ and $2m_K^2 - m_\pi^2 = (m_{K^+}^2 + m_{K^0}^2 - m_{\pi^+}^2)_{\mathrm{phys}}$ \cite{Urech:1994hd, Neufeld:1995mu}, which yields the physical masses  $m_\pi = 134.9768(5) \ \mathrm{MeV}$ and $m_K = 495.011(10)\ \mathrm{MeV}$ for the pion and kaon, respectively.
In practice, to describe the chiral dependence we employ  the dimensionless hadronic combinations
\begin{equation}
	\Phi_2 = 8t_0 m_\pi^2, \qquad \Phi_4 = 8t_0\left(m_K^2+ \frac{1}{2} m_\pi^2\right),
\end{equation}
where the  gradient flow scale $t_0/a^2$ \cite{Luscher:2010iy} is used for scale setting. Its  physical value, $t_0^{\mathrm{phys}} = 0.1449(7)\ \mathrm{fm}$, is adopted from the $\Xi^0$-baryon mass determination in \cite{RQCD:2022xux}.

We parametrise the lattice spacing dependence guided by Symanzik effective theory. Given the relative short-distance nature of the observables and their very precise determination,  higher-order cutoff effects  must be incorporated into our fits. Our most general fit ansatz reads
\begin{equation}
		\mathcal{O}(X_a) = \beta_2 X_a^2 + \beta_3X_a^3 + \beta_4X_a^4 + \delta_2X_a^2 \left(
		\Phi_2 - \Phi_2^{\mathrm{phys}}
		\right) + \delta_3X_a^3 \left(
		\Phi_2 - \Phi_2^{\mathrm{phys}}
		\right)
		+ \epsilon_2 X_a^2 \left(
		\Phi_4 - \Phi_4^{\mathrm{phys}}
		\right),
\end{equation}
where $X_a^2 = a^2/(8t_0)$. In practice, it is challenging to constrain  all  fit parameters simultaneously, therefore we explore various functional forms by selectively dropping one or more of the terms multiplied by the parameters $\beta_i, \delta_i, \epsilon_2$.   To model the chiral dependence of observable $\mathcal{O}$, we include a linear term in $\Phi_2$ along with a  higher-order correction,
\begin{equation}
	\mathcal{O}\big(\Phi_2\big) = \mathcal{O}\big(\Phi_2^{\mathrm{phys}}\big) + \gamma_1 \left(
	\Phi_2 - \Phi_2^{\mathrm{phys}} 
	\right)
	+ \gamma_2\left(
	f_{\chi}(\Phi_2) - f_{\chi}(\Phi_2^{\mathrm{phys}})
	\right),
\end{equation}
where $f_{\chi}\in \{ \Phi_2\log(\Phi_2); \Phi_2^2\}$. The strange quark mass dependence, governed by  $\Phi_4$, is included as a linear term 
Since $\Phi_4$ is close to its physical value on all ensembles, higher-order corrections are unnecessary.
To assess systematic uncertainties arising from the chiral-continuum extrapolations, we apply  cuts to the data sets,  either excluding the coarsest lattice spacing or by removing all ensembles with $m_\pi > 400 \ \mathrm{MeV}$. For the final estimate and systematic error analysis,  we perform a weighted model average following   \cite{Jay:2020jkz}, with weights assigned according to the Takeuchi Information Criterion (TIC) \cite{Frison:2023lwb}.

\begin{figure}
	\centering
	\includegraphics[scale=0.27]{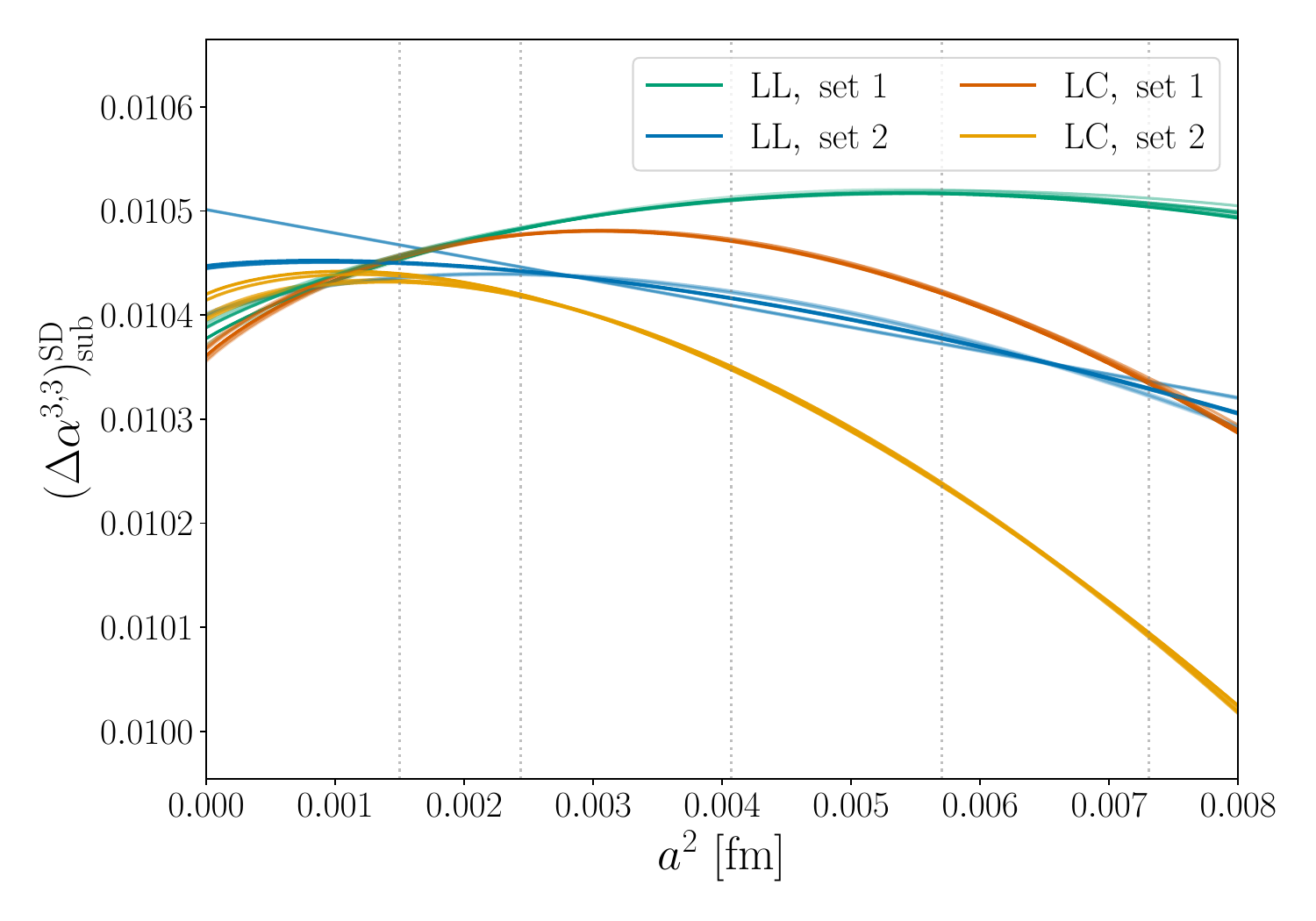}
	\includegraphics[scale=0.28]{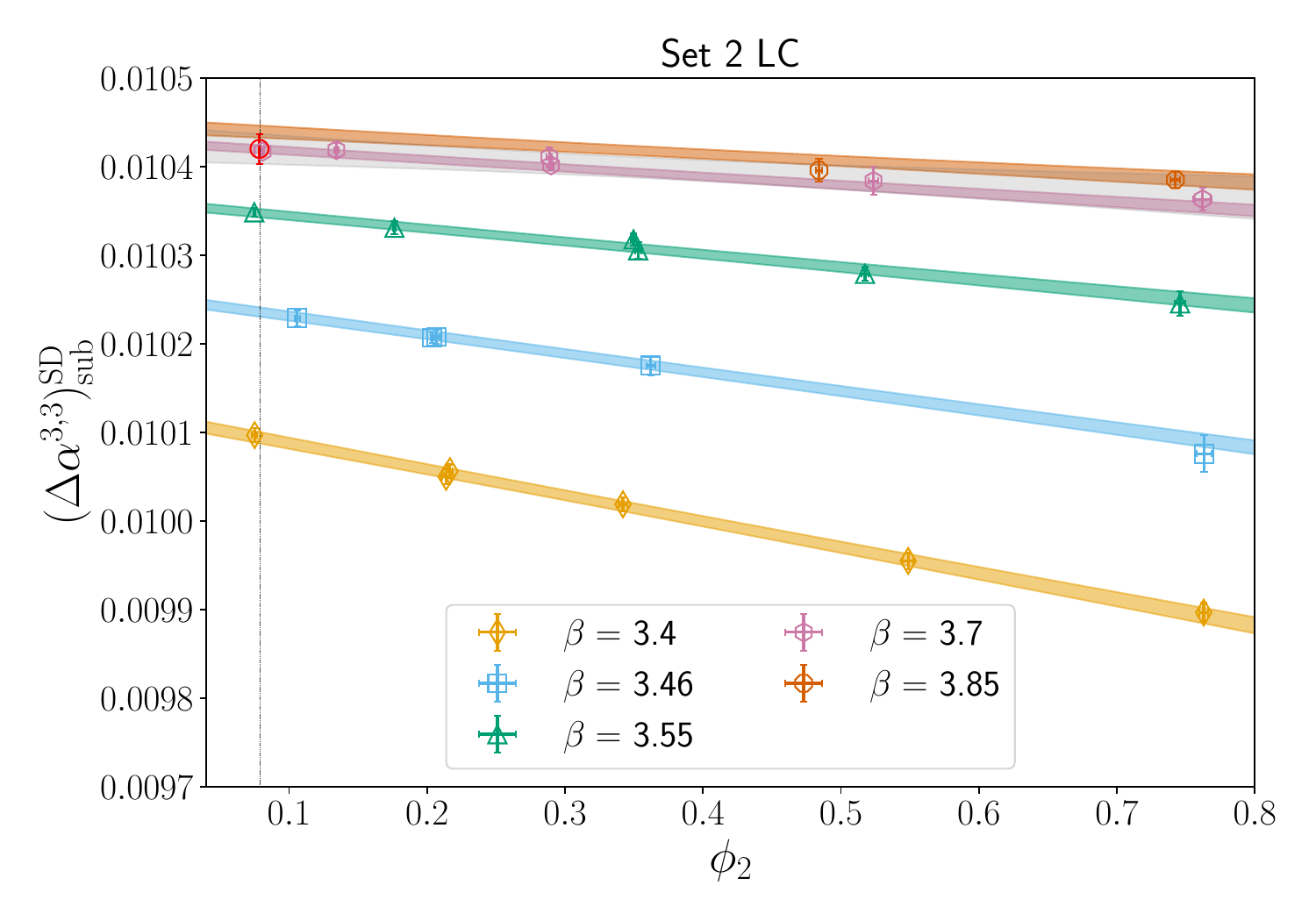}
	\caption{Illustration of fits to the isovector contribution in the Short Distance (SD) region. \textit{Left:} continuum limit behaviours for fours sets of data based on different improvement schemes and discretisations of the vector current. Each line corresponds to a single fit, with the opacity associated to the weights as given by our model average prescription.  \textit{Right:} chiral approach to the physical pion mass for one of the  fits with the highest weight. Data points are projected to $\Phi_4^{\mathrm{phys}}$. Coloured lines denote the chiral trajectories at finite lattice spacing, while the grey band shows the dependence on $\Phi_2$ in the continuum. Results are shown  for $Q^2=Q_m^2= 9\  \mathrm{GeV}^2$. }
	\label{fig:isovector_cl}
\end{figure}

On the left-hand side of Fig.~\ref{fig:isovector_cl}, we present the continuum limit approach of the isovector contribution at physical quark masses for all the explored models. The right-hand side illustrates the chiral dependence based on the  best fit results obtained  through  our model average prescription. 

\begin{figure}
	\centering
	\includegraphics[scale=0.27]{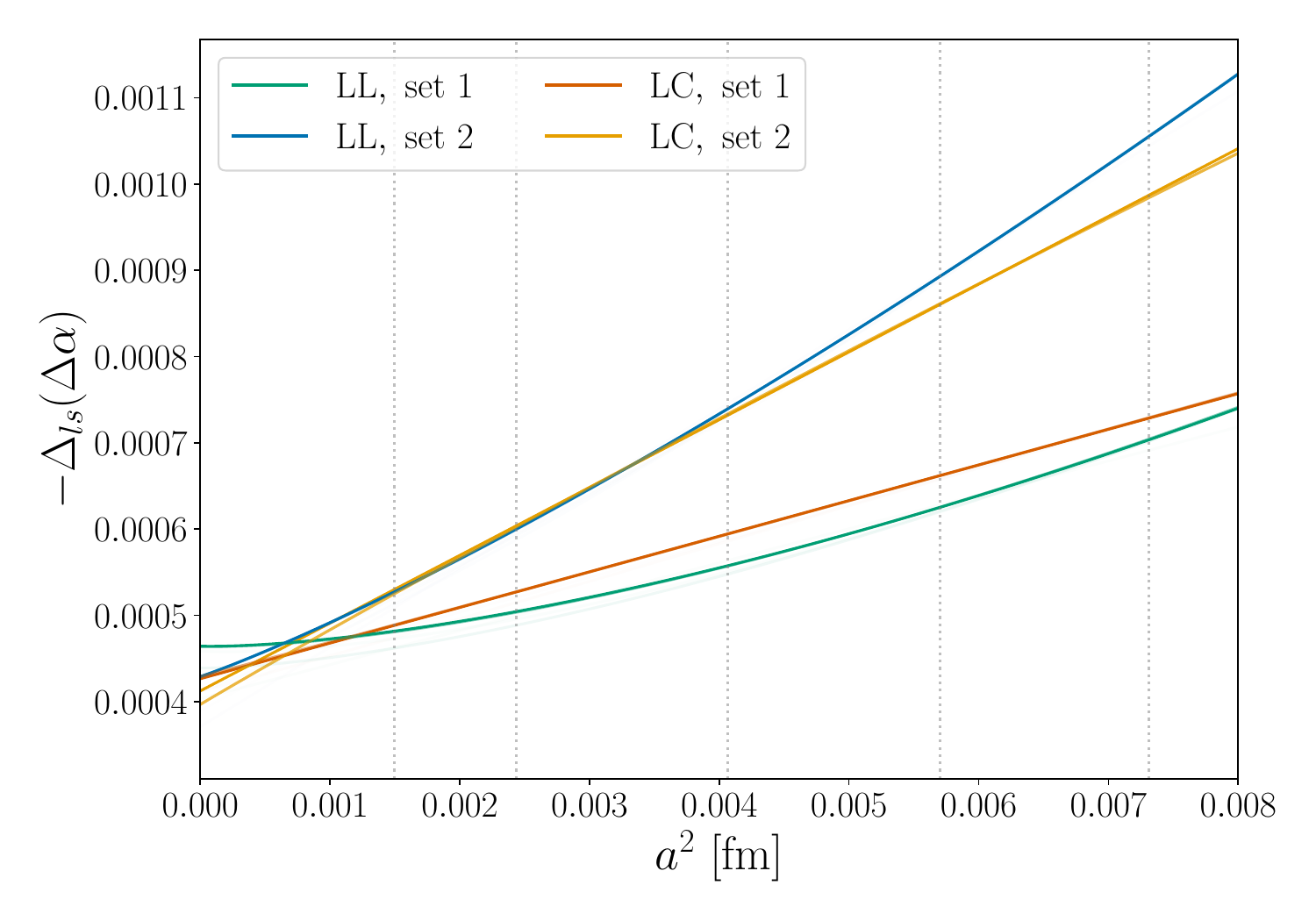}
	\includegraphics[scale=0.28]{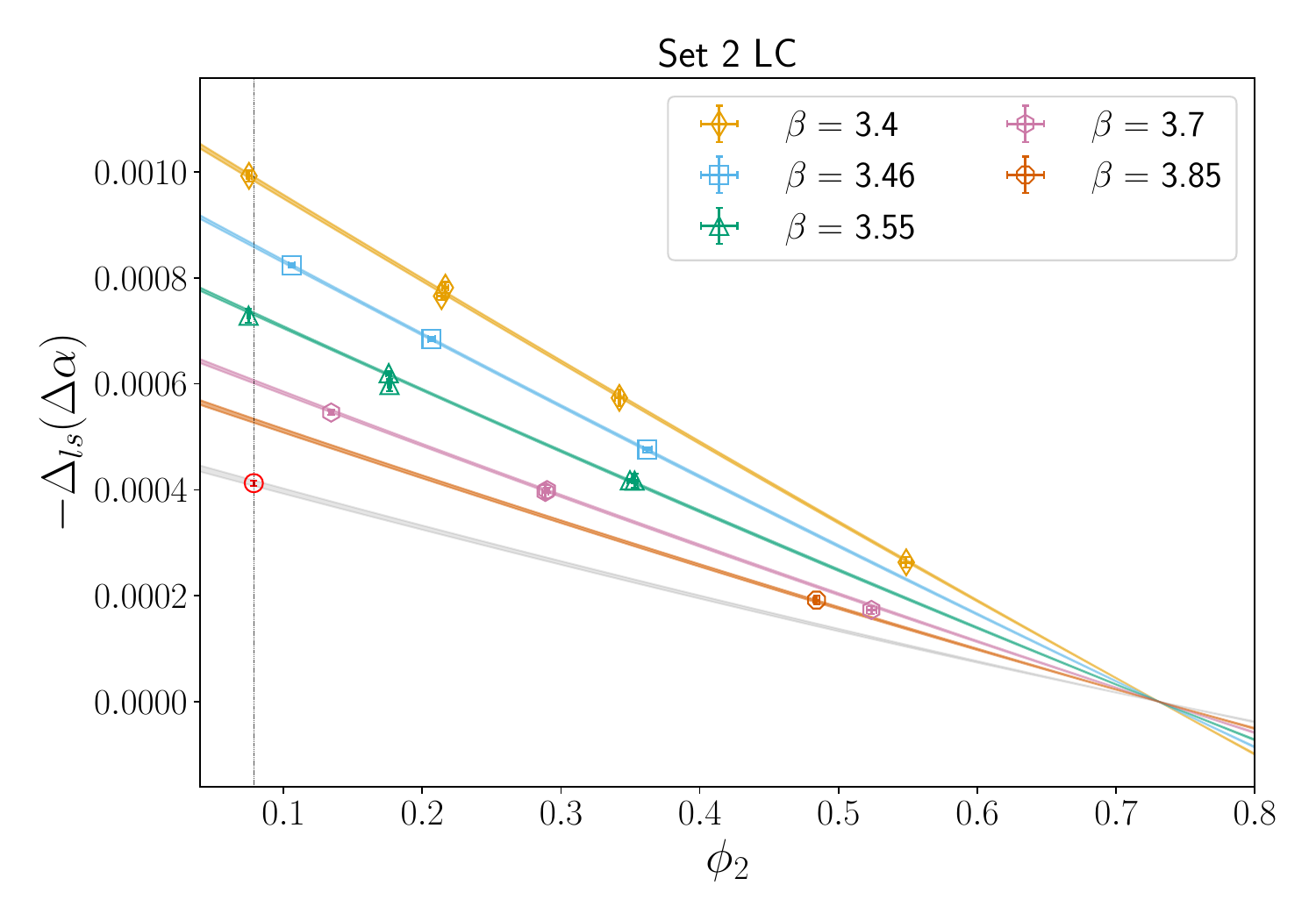}
	\caption{Illustration of fits to the $\Delta_{ls}$ contribution. \textit{Left:} continuum limit behaviours for four sets of data based on different improvement schemes and discretisations of the vector current.  \textit{Right:} chiral approach to the physical point for one of the best fits for a reference set.  Results are shown  for $Q^2=9\ \mathrm{GeV}^2$. }
	\label{fig:isoscalar_cl}
\end{figure}

The  isoscalar contribution is  computed by evaluating  $\Delta_{ls}(\Delta\alpha)$, as defined  in Eq.~(\ref{eq:isoscalar_decomposition}). This quantity vanishes at the $SU(3)$-symmetric point and at leading order scales proportionally to $m_s - m_l$. To capture this behaviour, we model the chiral-continuum dependence using the following parametrisation
\begin{equation}
	\Delta_{ls}(\Delta\alpha)(\Phi_\delta, \Phi_4, X_a) = \Phi_\delta\left(
	\gamma_1 + \gamma_2\Phi_\delta + \beta_2X_a^2 + \beta_3X_a^3 + \gamma_0 \Phi_4
	\right),
	\label{eq:su3_fit_ansatz}
\end{equation}
where $\Phi_\delta = \Phi_4 - \frac{3}{2}\Phi_2$. This formulation ensures that all cutoff effects are suppressed by $\Phi_\delta$ in the proximity of the $SU(3)$-symmetric point. An illustration of the continuum limit behaviour and the chiral-dependence for $\Delta_{ls}$ is shown in Fig.~\ref{fig:isoscalar_cl}. 
Similarly, the $\widehat{\mathit{\Pi}}^{(0,8)}$ contribution vanishes linearly as the $SU(3)$-symmetric point is approached.  We therefore describe the chiral-continuum dependence using  a  fit ansatz similar to  Eq.~(\ref{eq:su3_fit_ansatz}). 

Finally,  for the charm connected contribution we evaluate non-perturbatively both $\widehat{\mathit{\Pi}}^{(c,c)}_{\mathrm{sub}}$ and $\Delta_{lc}b$ as introduced in Eq.~(\ref{eq:charm_decomposition}). The charm quark mass tuning  to match the physical $D_s$ meson mass of $1968.47\ \mathrm{MeV}$ and the evaluation of the renormalization constant for the local vector current are described in \cite{Ce:2022kxy,Gerardin:2019rua}.  The extrapolation to the physical point follows the approach used for the isovector contribution, however we exclude higher-order corrections in $\Phi_2$, as we expect a mild dependence on the pion mass. For the final value we exclude  the local-local current discretization from the model average due to its significantly larger cutoff effects.


\section{Running with energy}
\begin{figure}
	\centering
	\includegraphics[scale=0.42]{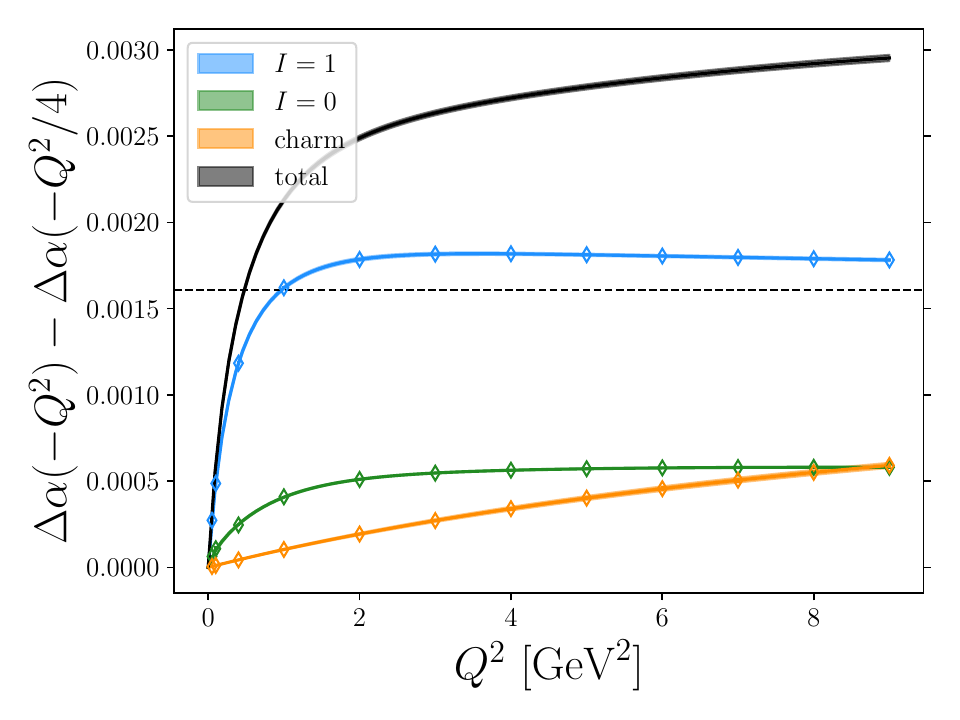}
	\includegraphics[scale=0.42]{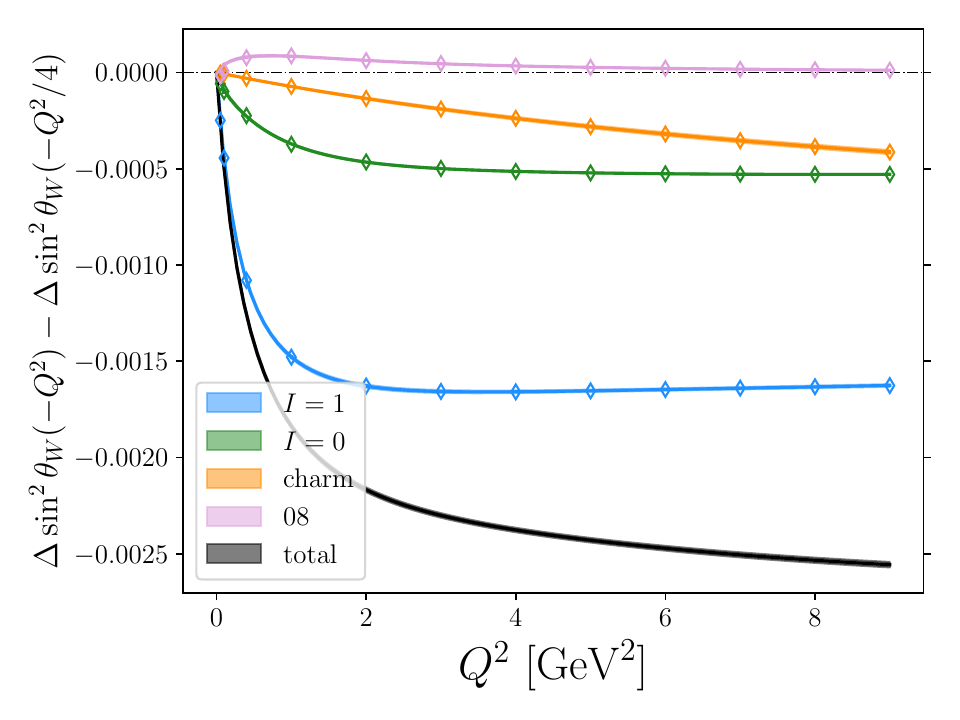}
	\caption{HVP contribution computed in the region $Q^2-Q^2/4$ to the running of $\alpha$ (\textit{left}) and $\sin^2\theta_W$ (\textit{right}) as a function of $Q^2$. Different colours represent the isovector ($\mathit{I}=1$), isoscalar ($\mathit{I}=0$), charm and, for $\sin^2\theta_W$, the mixed $Z\gamma$ contributions. The dashed line on the left panel denotes the tree-level value for the isovector contribution.  }
	\label{fig:ew_running}
\end{figure}

In our study, we sampled the HVP function at multiple values of $Q^2$ within the range  $0.1\  \mathrm{GeV}^2\leq Q^2 \leq 9 \ \mathrm{GeV}^2$. To provide a smooth and  analytic representation of the HVP momentum dependence  we employ a Pad\'{e} approximant, whose general expression reads \cite{Aubin2012,Ce:2022eix}
\begin{equation}
	\widehat{\mathit{\Pi}}(Q^2) \approx R_M^N(Q^2) = 
	\frac{\sum_{j=0}^{M} a_j Q^{2j} }
	{1 + \sum_{k=1}^{N}b_kQ^{2k}}.
\end{equation} 
Here, numerator and denominator are polynomials of degree $M$ and $N$, respectively. Since the subtracted HVP function vanishes at $Q^2=0$, we impose the constraint $a_0=0$ during the fitting process to ensure consistency.  We find that polynomials of  degree $M=2$  and $N=3$ accurately describe the data, while higher order coefficients are poorly determined. The  results of this analysis for the quantities $\Delta\alpha_{\mathrm{had}}$ and $\Delta_{\mathrm{had}} \sin^2\theta_W$ in the high-energy regime, specifically $Q^2-Q^2/4$, are illustrated in Fig.~\ref{fig:ew_running}.

\section{Conclusion}
We have presented preliminary results for the leading hadronic contribution to the running of the electromagnetic coupling and the electroweak mixing angle in the high-energy regime by computing the $\widehat{\mathit{\Pi}}(Q^2) = 	\mathit{\Pi}(Q^2) -  \mathit{\Pi}(Q^2/4)$  HVP  across the momentum  range $Q^2\leq 9 \ \mathrm{GeV}^2$. This decomposition allows us to perform  a clear separation of contributions from different Euclidean regions, thus giving access to higher values of the  space-like momenta compared to our 2022 study \cite{Ce:2022eix}. This method also provides improved control over the systematic uncertainties associated with the extrapolation to the physical point. Looking ahead, we plan to compute the missing low-energy component, $\mathit{\Pi}(Q^2/4) -  \mathit{\Pi}(0)$, within the same momentum range. Combining our lattice results with perturbative QCD, we aim to obtain a high-precision  estimate of $\Delta\alpha_{\mathrm{had}}^{(5)}(M_Z^2)$. Together with additional statistics and improved noise reduction techniques employed in the calculation of the vector correlator $G(t)$, we expect a significantly  improved accuracy for $\Delta\alpha_{\mathrm{had}}$ at the Z-pole  with respect to \cite{Ce:2022eix}. This improved precision will be crucial to corroborate the observed tension with data-driven approaches for $\Delta\alpha_{\mathrm{had}}(Q^2)$.
In addition, this will either further confirm or disprove the observed consistency between lattice results and global electroweak fits in the estimation of $\Delta\alpha_{\mathrm{had}}^{(5)}(M_Z^2)$.

\acknowledgments{\noindent
Calculations for this project have been performed on the HPC
clusters Clover and HIMster-II at Helmholtz Institute Mainz and
Mogon-II and Mogon-NHR at Johannes Gutenberg-Universität (JGU)
Mainz, on the HPC systems JUQUEEN and JUWELS and on the GCS
Supercomputers HAZELHEN and HAWK at Höchstleistungsrechenzentrum
Stuttgart (HLRS).
The authors gratefully acknowledge the support of the Gauss Centre
for Supercomputing (GCS) and the John von Neumann-Institut für
Computing (NIC) projects HMZ21, HMZ23 and HINTSPEC at JSC, as well
as projects GCS-HQCD and GCS-MCF300 at HLRS. We also gratefully
acknowledge the scientific support and HPC resources provided by
NHR-SW of Johannes Gutenberg-Universität Mainz (project NHR-Gitter).
This work has been supported by Deutsche Forschungsgemeinschaft
(German Research Foundation, DFG) through Project HI~2048/1-2
(Project No.\ 399400745) and through the Cluster of Excellence
``Precision Physics, Fundamental Interactions and Structure of
Matter'' (PRISMA+ EXC 2118/1), funded within the German Excellence
strategy (Project No.\ 39083149). This project has received funding
from the European Union's Horizon Europe research and innovation
programme under the Marie Sk\l{}odowska-Curie grant agreement
No.\ 101106243. }

\bibliographystyle{JHEP}
\bibliography{proceedingsLat24.bib}

\providecommand{\href}[2]{#2}\begingroup\raggedright\begin{thebibliography}{10}

\bibitem{Ce:2022eix}
M.~C\`e, A.~G\'erardin, G.~von Hippel, H.~B. Meyer, K.~Miura, K.~Ottnad et~al.,
  \emph{{The hadronic running of the electromagnetic coupling and the
  electroweak mixing angle from lattice QCD}},
  \href{http://dx.doi.org/10.1007/JHEP08(2022)220}{\emph{JHEP} {\bf 08} (2022)
  220}, [\href{http://arxiv.org/abs/2203.08676}{{\tt 2203.08676}}].

\bibitem{Muong-2:2021ojo}
\textsc{(Muon g-2)}, B.~Abi et~al., \emph{{Measurement of the Positive Muon
  Anomalous Magnetic Moment to 0.46 ppm}},
  \href{http://dx.doi.org/10.1103/PhysRevLett.126.141801}{\emph{Phys. Rev.
  Lett.} {\bf 126} (2021) 141801}, [\href{http://arxiv.org/abs/2104.03281}{{\tt
  2104.03281}}].

\bibitem{Muong-2:2023cdq}
\textsc{(Muon g-2)}, D.~P. Aguillard et~al., \emph{{Measurement of the Positive
  Muon Anomalous Magnetic Moment to 0.20~ppm}},
  \href{http://dx.doi.org/10.1103/PhysRevLett.131.161802}{\emph{Phys. Rev.
  Lett.} {\bf 131} (2023) 161802}, [\href{http://arxiv.org/abs/2308.06230}{{\tt
  2308.06230}}].

\bibitem{Borsanyi:2020mff}
S.~Borsanyi et~al., \emph{{Leading hadronic contribution to the muon magnetic
  moment from lattice QCD}},
  \href{http://dx.doi.org/10.1038/s41586-021-03418-1}{\emph{Nature} {\bf 593}
  (2021) 51}, [\href{http://arxiv.org/abs/2002.12347}{{\tt 2002.12347}}].

\bibitem{Djukanovic:2024cmq}
D.~Djukanovic, G.~von Hippel, S.~Kuberski, H.~B. Meyer, N.~Miller, K.~Ottnad
  et~al., \emph{{The hadronic vacuum polarization contribution to the muon
  $g-2$ at long distances}},  \href{http://arxiv.org/abs/2411.07969}{{\tt
  2411.07969}}.

\bibitem{Boccaletti:2024guq}
A.~Boccaletti et~al., \emph{{High precision calculation of the hadronic vacuum
  polarisation contribution to the muon anomaly}},
  \href{http://arxiv.org/abs/2407.10913}{{\tt 2407.10913}}.

\bibitem{RBC:2024fic}
\textsc{(RBC, UKQCD)}, T.~Blum et~al., \emph{{The long-distance window of the
  hadronic vacuum polarization for the muon g-2}},
  \href{http://arxiv.org/abs/2410.20590}{{\tt 2410.20590}}.

\bibitem{Bazavov:2024eou}
A.~Bazavov et~al., \emph{{Hadronic vacuum polarization for the muon $g-2$ from
  lattice QCD: Long-distance and full light-quark connected contribution}},
  \href{http://arxiv.org/abs/2412.18491}{{\tt 2412.18491}}.

\bibitem{Becker:2018ggl}
D.~Becker et~al., \emph{{The P2 experiment}},
  \href{http://dx.doi.org/10.1140/epja/i2018-12611-6}{\emph{Eur. Phys. J. A}
  {\bf 54} (2018) 208}, [\href{http://arxiv.org/abs/1802.04759}{{\tt
  1802.04759}}].

\bibitem{Bruno:2016plf}
M.~Bruno, T.~Korzec and S.~Schaefer, \emph{{Setting the scale for the CLS $2 +
  1$ flavor ensembles}},
  \href{http://dx.doi.org/10.1103/PhysRevD.95.074504}{\emph{Phys. Rev. D} {\bf
  95} (2017) 074504}, [\href{http://arxiv.org/abs/1608.08900}{{\tt
  1608.08900}}].

\bibitem{Mohler:2017wnb}
D.~Mohler, S.~Schaefer and J.~Simeth, \emph{{CLS 2+1 flavor simulations at
  physical light- and strange-quark masses}},
  \href{http://dx.doi.org/10.1051/epjconf/201817502010}{\emph{EPJ Web Conf.}
  {\bf 175} (2018) 02010}, [\href{http://arxiv.org/abs/1712.04884}{{\tt
  1712.04884}}].

\bibitem{Mohler:2020txx}
D.~Mohler and S.~Schaefer, \emph{{Remarks on strange-quark simulations with
  Wilson fermions}},
  \href{http://dx.doi.org/10.1103/PhysRevD.102.074506}{\emph{Phys. Rev. D} {\bf
  102} (2020) 074506}, [\href{http://arxiv.org/abs/2003.13359}{{\tt
  2003.13359}}].

\bibitem{Kuberski:2023zky}
S.~Kuberski, \emph{{Low-mode deflation for twisted-mass and RHMC reweighting in
  lattice QCD}},
  \href{http://dx.doi.org/10.1016/j.cpc.2024.109173}{\emph{Comput. Phys.
  Commun.} {\bf 300} (2024) 109173},
  [\href{http://arxiv.org/abs/2306.02385}{{\tt 2306.02385}}].

\bibitem{Bali:2016umi}
\textsc{(RQCD)}, G.~S. Bali, E.~E. Scholz, J.~Simeth and W.~S\"oldner,
  \emph{{Lattice simulations with $N_f=2+1$ improved Wilson fermions at a fixed
  strange quark mass}},
  \href{http://dx.doi.org/10.1103/PhysRevD.94.074501}{\emph{Phys. Rev. D} {\bf
  94} (2016) 074501}, [\href{http://arxiv.org/abs/1606.09039}{{\tt
  1606.09039}}].

\bibitem{Gerardin:2018kpy}
A.~Gerardin, T.~Harris and H.~B. Meyer, \emph{{Nonperturbative renormalization
  and $O(a)$-improvement of the nonsinglet vector current with $N_f=2+1$ Wilson
  fermions and tree-level Symanzik improved gauge action}},
  \href{http://dx.doi.org/10.1103/PhysRevD.99.014519}{\emph{Phys. Rev. D} {\bf
  99} (2019) 014519}, [\href{http://arxiv.org/abs/1811.08209}{{\tt
  1811.08209}}].

\bibitem{Heitger:2020zaq}
\textsc{(ALPHA)}, J.~Heitger and F.~Joswig, \emph{{The renormalised
  $\mathrm{O}(a)$ improved vector current in three-flavour lattice QCD with
  Wilson quarks}},
  \href{http://dx.doi.org/10.1140/epjc/s10052-021-09037-4}{\emph{Eur. Phys. J.
  C} {\bf 81} (2021) 254}, [\href{http://arxiv.org/abs/2010.09539}{{\tt
  2010.09539}}].

\bibitem{Fritzsch:2018zym}
P.~Fritzsch, \emph{{Mass-improvement of the vector current in three-flavor
  QCD}}, \href{http://dx.doi.org/10.1007/JHEP06(2018)015}{\emph{JHEP} {\bf 06}
  (2018) 015}, [\href{http://arxiv.org/abs/1805.07401}{{\tt 1805.07401}}].

\bibitem{Ce:2022kxy}
M.~C\`e et~al., \emph{{Window observable for the hadronic vacuum polarization
  contribution to the muon g-2 from lattice QCD}},
  \href{http://dx.doi.org/10.1103/PhysRevD.106.114502}{\emph{Phys. Rev. D} {\bf
  106} (2022) 114502}, [\href{http://arxiv.org/abs/2206.06582}{{\tt
  2206.06582}}].

\bibitem{Bernecker:2011gh}
D.~Bernecker and H.~B. Meyer, \emph{{Vector Correlators in Lattice QCD: Methods
  and applications}},
  \href{http://dx.doi.org/10.1140/epja/i2011-11148-6}{\emph{Eur. Phys. J. A}
  {\bf 47} (2011) 148}, [\href{http://arxiv.org/abs/1107.4388}{{\tt
  1107.4388}}].

\bibitem{Francis:2013fzp}
A.~Francis, B.~Jaeger, H.~B. Meyer and H.~Wittig, \emph{{A new representation
  of the Adler function for lattice QCD}},
  \href{http://dx.doi.org/10.1103/PhysRevD.88.054502}{\emph{Phys. Rev. D} {\bf
  88} (2013) 054502}, [\href{http://arxiv.org/abs/1306.2532}{{\tt 1306.2532}}].

\bibitem{Kuberski:2024bcj}
S.~Kuberski, M.~C\`e, G.~von Hippel, H.~B. Meyer, K.~Ottnad, A.~Risch et~al.,
  \emph{{Hadronic vacuum polarization in the muon g \ensuremath{-} 2: the
  short-distance contribution from lattice QCD}},
  \href{http://dx.doi.org/10.1007/JHEP03(2024)172}{\emph{JHEP} {\bf 03} (2024)
  172}, [\href{http://arxiv.org/abs/2401.11895}{{\tt 2401.11895}}].

\bibitem{Ce:2021xgd}
M.~C\`e, T.~Harris, H.~B. Meyer, A.~Toniato and C.~T\"or\"ok, \emph{{Vacuum
  correlators at short distances from lattice QCD}},
  \href{http://dx.doi.org/10.1007/JHEP12(2021)215}{\emph{JHEP} {\bf 12} (2021)
  215}, [\href{http://arxiv.org/abs/2106.15293}{{\tt 2106.15293}}].

\bibitem{Sommer:2022wac}
R.~Sommer, L.~Chimirri and N.~Husung, \emph{{Log-enhanced discretization errors
  in integrated correlation functions}},
  \href{http://dx.doi.org/10.22323/1.430.0358}{\emph{PoS} {\bf LATTICE2022}
  (2023) 358}, [\href{http://arxiv.org/abs/2211.15750}{{\tt 2211.15750}}].

\bibitem{Urech:1994hd}
R.~Urech, \emph{{Virtual photons in chiral perturbation theory}},
  \href{http://dx.doi.org/10.1016/0550-3213(95)90707-N}{\emph{Nucl. Phys. B}
  {\bf 433} (1995) 234}, [\href{http://arxiv.org/abs/hep-ph/9405341}{{\tt
  hep-ph/9405341}}].

\bibitem{Neufeld:1995mu}
H.~Neufeld and H.~Rupertsberger, \emph{{The Electromagnetic interaction in
  chiral perturbation theory}},
  \href{http://dx.doi.org/10.1007/s002880050156}{\emph{Z. Phys. C} {\bf 71}
  (1996) 131}, [\href{http://arxiv.org/abs/hep-ph/9506448}{{\tt
  hep-ph/9506448}}].

\bibitem{Luscher:2010iy}
M.~L{\"u}scher, \emph{{Properties and uses of the Wilson flow in lattice QCD}},
  \href{http://dx.doi.org/10.1007/JHEP08(2010)071}{\emph{JHEP} {\bf 08} (2010)
  071}, [\href{http://arxiv.org/abs/1006.4518}{{\tt 1006.4518}}]. [Erratum:
  \href{https://doi.org/10.1007/JHEP03(2014)092}{JHEP \textbf{03} (2014) 092}].

\bibitem{RQCD:2022xux}
\textsc{(RQCD)}, G.~S. Bali, S.~Collins, P.~Georg, D.~Jenkins, P.~Korcyl,
  A.~Sch\"afer et~al., \emph{{Scale setting and the light baryon spectrum in
  N$_{f}$ = 2 + 1 QCD with Wilson fermions}},
  \href{http://dx.doi.org/10.1007/JHEP05(2023)035}{\emph{JHEP} {\bf 05} (2023)
  035}, [\href{http://arxiv.org/abs/2211.03744}{{\tt 2211.03744}}].

\bibitem{Jay:2020jkz}
W.~I. Jay and E.~T. Neil, \emph{{Bayesian model averaging for analysis of
  lattice field theory results}},
  \href{http://dx.doi.org/10.1103/PhysRevD.103.114502}{\emph{Phys. Rev. D} {\bf
  103} (2021) 114502}, [\href{http://arxiv.org/abs/2008.01069}{{\tt
  2008.01069}}].

\bibitem{Frison:2023lwb}
J.~Frison, \emph{{Towards fully bayesian analyses in Lattice QCD}},
  \href{http://arxiv.org/abs/2302.06550}{{\tt 2302.06550}}.

\bibitem{Gerardin:2019rua}
A.~G\'erardin, M.~C\`e, G.~von Hippel, B.~H\"orz, H.~B. Meyer, D.~Mohler
  et~al., \emph{{The leading hadronic contribution to $(g-2)_\mu$ from lattice
  QCD with $N_{\rm f}=2+1$ flavours of O($a$) improved Wilson quarks}},
  \href{http://dx.doi.org/10.1103/PhysRevD.100.014510}{\emph{Phys. Rev. D} {\bf
  100} (2019) 014510}, [\href{http://arxiv.org/abs/1904.03120}{{\tt
  1904.03120}}].

\bibitem{Aubin2012}
C.~Aubin, T.~Blum, M.~Golterman and S.~Peris, \emph{Model-independent
  parametrization of the hadronic vacuum polarization and $g$-2 for the muon on
  the lattice}, \href{http://dx.doi.org/10.1103/PhysRevD.86.054509}{\emph{Phys.
  Rev. D} {\bf 86} (2012) 054509}, [\href{http://arxiv.org/abs/1205.3695}{{\tt
  1205.3695}}].

\end{thebibliography}\endgroup

\end{document}